\documentclass[14pt]{article}
\usepackage[utf8]{inputenc}
\usepackage{cite}
\usepackage{authblk}
\usepackage[dvipsnames]{xcolor}
\usepackage{graphicx}
\usepackage{subfig}
\usepackage[margin=1.0in]{geometry}
\usepackage{amsmath}
\usepackage{amssymb}
\usepackage{physics}
\usepackage{cancel}
\usepackage{comment}
\usepackage{tikz}
\usepackage{mwe}
\usepackage{mathtools}
\definecolor{electricblue}{RGB}{0,50,255}
\RequirePackage[colorlinks,citecolor=magenta,urlcolor=magenta,linkcolor=magenta]{hyperref}
\allowdisplaybreaks
\labelformat{section}{Section #1} 
\labelformat{subsection}{Section #1} 
\labelformat{subsubsection}{Section #1}
\labelformat{subsubsubsection}{Section #1}
\labelformat{equation}{Eq.~(#1)} 
\labelformat{subfigure}{Fig.~\thefigure#1} \labelformat{table}{Table~#1} 
\labelformat{appendix}{Appendix #1}

\usepackage{geometry}
\usepackage{hyperref}
\definecolor{darkslateblue}{rgb}{0.28, 0.24, 0.55}

\newcounter{casebrane}

\newcounter{subcasebrane}[casebrane]

\newcounter{caseLovelock}

\makeatletter
\def\@fnsymbol#1{\ensuremath{\ifcase#1\or \ddagger\or \dagger\or
   \mathsection\or \mathparagraph\or \|\or **\or \ddagger\ddagger
   \or \dagger\dagger \else\@ctrerr\fi}}

\title{Maximal force bounds in general relativity and beyond}
\author{Sanhitaa Guha\footnote{
\href{mailto:sanhitaag@imsc.res.in}{sanhitaag@imsc.res.in}}}
\affil{{\small{{{The Institute of Mathematical Sciences, CIT Campus, Chennai 600113, India}}}}}
\date{}

\begin{document}

\maketitle


\begin{abstract}
  Gibbons and Schiller proposed the maximal force conjecture in general relativity, according to which the maximum force is bounded by $c^4/G$. In this work, we derive upper bounds on the force for stable, self-gravitating, spatially regular and horizonless compact objects in $d$-dimensional general relativity, braneworld gravity, pure Lovelock gravity and Einstein–Gauss–Bonnet gravity. Our analysis uses the connection between photon-ring structure and the stability of horizonless compact objects, adopting the absence of light rings as a criterion for stability. We investigate how the resulting force bounds depend on the spacetime dimension and on modifications to gravitational dynamics arising from extra dimensions and higher curvature corrections. We study how the bounds for such systems change under alternative higher curvature theories of gravity or in higher dimensions. Throughout this work, we consider a static and spherically symmetric spacetime sourced by anisotropic matter for which the radial and tangential pressures may differ.
\end{abstract}
\

\section{Introduction}
\label{sec:introduction}
The detection of gravitational waves from compact object mergers and the imaging of black hole shadows have provided compelling observational support for the black-hole paradigm. Nevertheless, the possibility that horizonless compact objects may reproduce some of the exterior gravitational properties of black holes has motivated considerable interest in alternative compact configurations. Among these, exotic compact objects (ECOs) constitute a broad class of proposed black-hole alternatives, including models such as gravastars and boson stars \cite{Kaup:1968zz, Ruffini:1969qy, Mazur:2001fv, Visser:2003ge, Cardoso:2019rvt}. These configurations provide a setting for investigating the properties of highly compact gravitational systems without event horizons and, in suitable models, with regular and singularity-free interiors. Their physical viability, however, depends on satisfying appropriate conditions of regularity and stability. This motivates the study of general constraints for stability of such regular, static, spherically symmetric, horizonless self-gravitating configurations.\par

Gibbons \cite{Gibbons:2002iv}  and Schiller \cite{Schiller:2004motion,Schiller:2021jxf} proposed a maximum-force conjecture according to which the force in general relativity cannot exceed a characteristic scale set by the fundamental constants,
\begin{equation}
F\leq \eta\frac{c^4}{G},
\label{eqn:gibb-schiller}
\end{equation}
where $\eta$ is a dimensionless constant. The strong form of the conjecture corresponds to $\eta=1/4$, whereas its weaker form allows $\eta=\mathcal{O}(1)$. This conjecture provides a useful motivation for examining whether analogous force bounds can be established for self-gravitating matter configurations subject to regularity, horizonlessness, and appropriate stability requirements. \par

An important aspect of the study of ultracompact objects (UCOs) is the existence and stability of circular null geodesics commonly referred to as photon rings. From the theorem in \cite{Cunha:2017qtt}, any axisymmetric and stationary solution to the Einstein field equations which results from the classical gravitational collapse of matter that satisfies the null energy condition and is smooth everywhere and ultra-compact (meaning it possesses at least one circular light orbit) must contain at least two light rings, with at least one (innermost) being stable \cite{Hod:2014ena,Hod:2017zpi}. Under general assumptions, a UCO formed from regular, nearly flat spacetime conditions must contain at least two light rings — one corresponding to a saddle point and the other to an extremum in an effective potential. Depending on the specifics, the extremum can represent a stable or unstable photon orbit. However, general relativity implies that an unstable configuration can only exist if the null energy condition is violated, which is a condition we assume to hold. Therefore, any physically reasonable UCO must possess at least one stable light ring. However, the existence of a stable light ring is problematic as it can trap massless field perturbations, leading to non-linear spacetime instabilities. These trapped disturbances can accumulate over time, eventually altering the spacetime through back-reaction and potentially leading to an instability \cite{Cunha:2022gde, Cunha:2025oeu, Cardoso:2014sna, Benomio:2018ivy, Keir:2014oka}. Thus, the matter configurations considered in the present work must be devoid of circular null geodesics to ensure their dynamical stability \cite{Hod:2024avc}. These factors contribute to a force-bound/re-scaled force bound (as we shall define later) for such stable self-gravitating horizonless matter configurations.

In this work, we investigate force bounds for stable, static, spherically symmetric, regular and horizonless self-gravitating configurations. We formulate the relevant force and rescaled-force quantities and derive constraints on them under the assumptions specified in our analysis. We then extend the investigation beyond $d$-dimensional general relativity to braneworld gravity and higher-curvature gravitational theories. Such extensions provide a setting in which to examine how modifications to the gravitational field equations affect the bounds obtained in general relativity and whether the resulting constraints retain a universal form or acquire dependence on the spacetime dimension, the coupling parameters or the properties of the matter distribution. \par

\textit{Notations and Conventions:} For deriving the force bounds, we adopt natural units by setting the fundamental constants $c$, $G$ and $\hbar$ to unity and use the mostly positive signature convention such that the flat spacetime metric takes the form $\eta_{\mu\nu}=\mathrm{diag}(-1,+1,+1,\ldots)$.
 
\section{Description of the system}
\label{sec:descriptionsystem}
We consider a stable, self-gravitating, spatially regular and horizonless matter configuration sourced by an anisotropic perfect fluid under the dominant energy condition. The spacetime geometry is described by the following static and spherically symmetric line element,\\
\begin{equation}\label{eqn:metric}
    ds^2=-e^{\nu(r)}dt^2+e^{\lambda(r)}dr^2+r^2d\Omega_{d-2}^2~. 
\end{equation}
Mathematically, the properties of the system under study translate to: 
\begin{enumerate}
\item{\textit{Spatial regularity}: This is characterized by the boundary condition \cite{Hod:2011aa,Hod:2011cxc,Hod:2013jhd},
\begin{equation}
\label{eqn:regular}
\begin{split}
    e^{-\lambda}(r\rightarrow 0)=1~~, ~~
    \nu(r \rightarrow0)<\infty~.
\end{split}
\end{equation}}
\item{\textit{Horizonlessness}: This is characterized by the condition that $e^{-\lambda (r)}\neq 0$, $\forall r\in[0,\infty)$. Thus,
\begin{equation}\label{eqn:horizonless}
  e^{-\lambda(r)}>0 ~,\quad \forall r\in[0,\infty)~. 
\end{equation}
}
Equivalently, the condition can be expressed as
\begin{equation}
\label{eqn:lambda_finite}
\lambda(r)<\infty~,
\quad \forall\,r\in[0,\infty)~.
\end{equation}
Working with an anisotropic perfect fluid, the components of the stress-energy-momentum tensor are given by $T^{tt}=\rho$, $T^{rr}=p$, $T^{\theta\theta}=p_T$ where $p $ is the radial pressure, $p_T$ is the transverse pressure, and $\rho$ is the density. Also, we consider the dominant energy condition to hold, which is given by
\begin{equation}\label{eqn:dominant}
    0\leq |p| \leq \rho~.
\end{equation}
\end{enumerate}
For the metric specified in \ref{eqn:metric}, the condition for circular null geodesics reduces to $r\nu'=2$, which can be expressed in the more general form $r g_{00}'=2g_{00}$. Thereby, we can define a function $\mathcal{N}(r)$ \cite{Chakraborty:2021dmu} from this condition such that it vanishes at the photon rings and acquires a positive value as $r \rightarrow \infty$. Thus, the condition for the absence of light rings, which is necessary for stability \cite{Hod:2024avc} as discussed before, is equivalent to
\begin{equation}\label{eqn:maincondition}
  \mathcal{N}(r)>0 ~, ~~\text{~~ $\forall r\in[0,\infty)$}~,
\end{equation}
provided that $\mathcal{N}(r)>0$ can be established at some $r$.

\section{Force in $d$-dimensional spacetime}
\label{sec:dforce}
In four-dimensional spacetime, a force function can be defined in terms of the radial pressure as $F=4\pi r^2p(r)$. In this work, we extend this definition to $d$-dimensional spacetime by adopting the following form,
\begin{equation}
\label{eqn:d-force}
F(r)=4\pi r^{d-2}p(r)~,
\end{equation}
where $p(r)$ denotes the radial pressure and $r$ is the radial coordinate. The factor $4\pi$ is retained in this definition for arbitrary spacetime dimensions.
The dimensional analysis of the $d$-dimensional force function can be expressed in terms of the reduced Planck constant $\hbar$, the speed of light $c$, and the $d$-dimensional gravitational constant $G_d$ as
\begin{equation}\label{eqn:dimanalysis}
   \displaystyle{[F_d]=[\hbar]^{\frac{d-4}{d-2}}[c]^{\frac{d+4}{d-2}}[G_d]^{\frac{-2}{d-2}}~}.
\end{equation}
For $d=4$, there is no dependence on $\hbar$ as
\begin{equation}\label{eqn:4dforceunit}
   [F_4]=\frac{c^4}{G_4}~.
\end{equation}
For $d \neq 4$, the expression is not independent of $\hbar$. However, for $F/r^{d-4}$, the dimensions are given by ${c^4}/{G_d}$. Thus, this rescaled force is independent of $\hbar$ \cite{Barrow:2014cga}. In higher-dimensional theories, when using natural units, the gravitational force is often rescaled to make it dimensionless or comparable across different spacetime dimensions or units. We shall see in further sections that in some cases we generally find bounds only for the rescaled force. 

We note that an alternative convention for the force function in $d$-dimensional spacetime is
\begin{equation} \label{eqn:d-force-omega}
F_{\Omega}(r)=\Omega_{d-2}r^{d-2}p(r)~,
\end{equation}
where
\begin{equation} \label{eqn:d-solid-angle}
\Omega_{d-2} = \frac{2\pi^{(d-1)/2}}{\Gamma\left(\frac{d-1}{2}\right)}
\end{equation} 
is the surface area of the unit $(d-2)$-sphere. This convention replaces the factor $4\pi$ in \ref{eqn:d-force} with $\Omega_{d-2}$. Consequently, the
two definitions are related by
\begin{equation} \label{eqn:force-conventions-relation}
F_{\Omega}(r)=\frac{\Omega_{d-2}}{4\pi}F(r).
\end{equation}
In four-dimensional spacetime, $\Omega_2=4\pi$, and the two definitions coincide. In this work, we retain the convention
in \ref{eqn:d-force} throughout.

We also note that the subsequent analysis is carried out in natural units, where $\hbar=c=G_d=1$. Therefore, to restore the physical dimensions of the force bounds obtained in $d$-dimensional spacetime, the corresponding bounds must be multiplied by the appropriate dimensional factor given in \ref{eqn:dimanalysis}. For the rescaled force $F_d/r^{d-4}$, this factor is $c^4/G_d$, which reduces to $c^4/G_4$ in four dimensions.

\section{Maximal force bounds}
\label{sec:maxforce}
In this section, we investigate upper bounds on the force function and its rescaled counterpart for our system in different gravitational theories. We begin with $d$-dimensional general relativity to establish a baseline for comparison before examining the effects of higher-dimensional bulk contributions in braneworld gravity and higher-curvature corrections in pure Lovelock and Einstein--Gauss--Bonnet gravity. We aim to determine how the spacetime dimension, the gravitational couplings, and the matter content influence the resulting force bounds.

\subsection{{$d$-dimensional general relativity}}
\label{subsec:d-gr}
We begin by deriving the force bound in $d$-dimensional general relativity, which provides a reference point. We then compare the higher-dimensional result with its four-dimensional counterpart. Using Einstein's field equations in $d$-dimensional spacetime \cite{Burikham:2015nma, Chakraborty:2021dmu}, the expression for $\lambda (r)$ and $\nu(r)$ is given by
\begin{equation}\label{eqn:gr1}
    r\lambda'e^{-\lambda}+(d-3)(1-e^{-\lambda})=(8\pi\rho+\Lambda)r^2~,
\end{equation}
\begin{equation}\label{eqn:gr2}
 r\nu'e^{-\lambda}-(d-3)(1-e^{-\lambda})=(8\pi p-\Lambda)r^2~,  
\end{equation}
where `prime' denotes the derivative with respect to radial coordinate $r$.\\
Solving the differential equation \ref{eqn:gr1} and integrating, we arrive at an explicit expression for $\lambda(r)$ given by
\begin{equation}\label{eqn:gr-lambda}
    e^{-\lambda}=1-\frac{2m(r)}{r^{d-3}}-\frac{\Lambda}{(d-1)}r^2~,
\end{equation}
where
\begin{equation}\label{eqn:gr-mass}
    m(r)=4\pi\int_0^{r}dr'\rho(r')r'^{d-2}~.
\end{equation}
Here, we note that we can integrate the density function from $0$ to infinity. This is appropriate for a horizonless, regular and spherically symmetric configuration. In contrast, for black-hole spacetimes, the presence of an event horizon and the associated singularity necessitate a different treatment, with the mass function typically defined by integrating outward from the horizon radius $r_{\rm H}$ and accounting separately for the mass enclosed within it.

Using the photon orbit condition $r\nu'=2$ in  \ref{eqn:gr2}, we obtain a function $\mathcal{N}_{\mathrm{GR},d}$ which vanishes on the light ring, that is, if $r_{\rm ph}$ is the photon circular orbit, $\mathcal{N}_{\mathrm{GR},d}(r_{\rm ph})=0$ \cite{Chakraborty:2021dmu}. The function is given by
\begin{equation}\label{eqn:Ngr1}
    \mathcal{N}_{\mathrm{GR},d}(r)=-8\pi pr^2+\Lambda r^2-(d-3)+(d-1)e^{-\lambda}~.
\end{equation}
Using the expression for $e^{- \lambda(r)}$ in terms of the mass variable $m(r)$ in  \ref{eqn:gr-lambda} in \ref{eqn:Ngr1}, we obtain
\begin{equation}\label{eqn:Ngr2}
    \mathcal{N}_{\mathrm{GR},d}(r)=2-2(d-1)\frac{m(r)}{r^{d-3}}-8\pi pr^2~.
\end{equation}
We see that $\mathcal{N}_{\mathrm{GR},d}(r)$ is independent of the cosmological constant $\Lambda$. It is further assumed that both the energy density $\rho(r)$ and pressure $p(r)$ decay sufficiently fast so that $pr^2 \to0$ and $m(r) \to$ constant as $r \to \infty$. From \ref{eqn:Ngr2}, it follows that
\begin{equation}\label{eqn:Ngrinf}
    \mathcal{N}_{\mathrm{GR},d}(r\to\infty)=2~.
\end{equation}

Thus, for a stable configuration, $\mathcal{N}_{\mathrm{GR},d}$ has to take a positive value throughout in order to account for the absence of null geodesics \cite{Hod:2024avc}. Using this condition from \ref{eqn:maincondition}, we obtain
\begin{equation}\label{eqn:gr-ineq1}
    4\pi pr^{d-2}<r^{d-4}-\frac{(d-1)m}{r}~.
\end{equation}
Since the dominant energy condition implies $\rho(r)\geq 0$, we have $m(r)\geq 0$. Consequently, the preceding inequality simplifies to
\begin{equation}\label{eqn:gr-ineq2}
    \frac{F_{\mathrm{GR},d}}{r^{d-4}}<\frac{c^4}{G_d}~,
\end{equation}
where $F_{\mathrm{GR},d}=4\pi pr^{d-2}$. We observe that we obtain a bound for $F_{\mathrm{GR},d}/r^{d-4}$, which is nothing but the rescaled force having the same dimensions as $c^4/G_d$. We define the rescaled force as
\begin{equation}\label{eqn:gr-rescaledf}
    \mathcal{F}_{\mathrm{GR},d}=\frac{F_{\mathrm{GR},d}}{r^{d-4}}~.
\end{equation}
With proper dimensions, the bound for the rescaled force is obtained as
\begin{equation}\label{eqn:gr-ineq3}
    \mathcal{F}_{\mathrm{GR},d}<\frac{c^4}{G_d}~.
\end{equation}
We can conclude that the force bound for $d<4$ decreases with an increase in radial distance $ r$, and for $d>4$, increases with an increase in $r$.
For $d=4$, $\mathcal{F}_{\mathrm{GR},4}=F_{\mathrm{GR},4}$ and we obtain
\begin{equation}\label{eqn:4dgr-ineq}
    F_{\mathrm{GR},4}<\frac{c^4}{G_4}~,
\end{equation}
which is in good agreement with the results that had been obtained in \cite{Hod:2024avc}.

\subsection{{Braneworld gravity}}
\label{subsec:braneworld}
In the previous section, we established an upper limit on the force by considering the effects of a cosmological constant and extra spatial dimensions in general relativity. In this section, we shift our focus to a similar higher-dimensional setup but now within the scope of effective gravitational field theories as observed by an observer confined to a four-dimensional brane. The brane will inherit additional correction terms in the effective field equation due to the presence of a higher five-dimensional bulk \cite{Shiromizu:1999wj}. Our goal in the following discussion is to explore how these extra contributions impact the previously derived bound.
The radial and temporal parts of the field equations with an effective four-dimensional brane cosmological constant $\Lambda_4$ are given by \cite{Chakraborty:2021dmu}
\begin{equation}\label{eqn:brane1}
    e^{-\lambda(r)}\left(\frac{1}{r^2}-\frac{\lambda'}{r}\right)-\frac{1}{r^2}=-8\pi G_4 \rho_{\rm eff}-\Lambda_4~,
\end{equation}
\begin{equation}\label{eqn:brane2}
    e^{-\lambda(r)}\left(\frac{1}{r^2}+\frac{\nu'}{r}\right)-\frac{1}{r^2}=8\pi G_4 p_{\rm eff}-\Lambda_4~.
\end{equation}
Here, we see that the result is structurally similar to that obtained in the previous section for $d=4$. However, here the physical density and pressure are replaced by effective density $\rho_{\rm eff}$ and effective pressure $p_{\rm eff}$ that incorporate quadratic matter corrections and contributions from the projected bulk Weyl tensor\cite{Maartens:2010ar, Chakraborty:2021dmu} and are given by
\begin{equation}\label{eqn:effdensity}
   \rho_{\rm eff}=\rho\left(1+\frac{\rho}{2 \lambda_{\rm b}}\right)+3\widetilde{U}~, 
\end{equation}
\begin{equation}\label{eqn:effpress}
   p_{\rm eff}=\left(p+\frac{\rho}{2\lambda_{\rm b}}(\rho+2p)\right)+\widetilde{U}+2\widetilde{P}~,  
\end{equation}
\begin{equation}\label{eqn:U_tilde}
    \widetilde{U}=2G_4U(8 \pi G_4)^{-2}\lambda_{\rm b}^{-1}~,
\end{equation}
\begin{equation}\label{eqn:P_tilde}
    \widetilde{P}=2G_4P(8 \pi G_4)^{-2}\lambda_{\rm b}^{-1}~,
\end{equation}
where, $U=-(G_4/G_5)^2E_{\mu\nu}u^\mu u^\nu$ and $P=(G_4/G_5)^2E_{\mu\nu}r^\mu r^\nu$. Here, $E_{\mu\nu}$ is the projected Weyl tensor on the bulk, $u^\mu$ is the four-velocity of a static observer, and $r_\mu$ is orthogonal to the four-velocity. $\lambda_{\rm b}$ is the brane tension, $G_4$ and $G_5$ are the four and five-dimensional gravitational constants, respectively, $\Lambda_4$ is the brane cosmological constant, while $\rho$ and $p$ correspond to the physical density and pressure due to the matter contribution, respectively. $\widetilde{U}$ and $\widetilde{P}$ are the dark radiation and the dark pressure arising from the effect of the higher-dimensional bulk. Since the equations are similar to the previous case, the solutions are also similar, with $p$ and $\rho$ being replaced by $p_{\rm eff}$ and $\rho_{\rm eff}$ and $d=4$ respectively. Hence, the mass term $m(r)$ here consists of the effective density and not only the matter density. Solving and integrating \ref{eqn:brane1} we obtain
\begin{equation}\label{eqn:brane-lambda}
    e^{-\lambda}=1-\frac{2m(r)}{r}-\frac{\Lambda_4}{3}r^2~,
\end{equation}
\begin{equation}\label{eqn:brane-mass}
    m(r)=4\pi\int_0^{r}dr'\rho_{\rm eff}(r')r'^{2}~.
\end{equation}
Similar to the previous case, we define $\mathcal{N}_{\rm brane}$ such that $\mathcal{N}_{\rm brane}(r_{\rm ph})=0$ using the photon orbit condition $r\nu'=2$ in \ref{eqn:brane2} \cite{Chakraborty:2021dmu}. Thus,
\begin{equation}\label{eqn:Nbrane1}
    \mathcal{N}_{\rm brane}(r)=-8\pi p_{\rm eff}r^2+\Lambda_4 r^2-1+3e^{-\lambda}~.
\end{equation}
Substituting  \ref{eqn:brane-lambda} in  \ref{eqn:Nbrane1} we get
\begin{equation}\label{eqn:Nbrane2}
    \mathcal{N}_{\rm brane}(r)=2-\frac{6m(r)}{r}-8\pi p_{\rm eff}r^2~.
\end{equation}
Thus, $\mathcal{N}_{\rm brane}$ is independent of $\Lambda_4$. We assume that $\rho_{\rm eff}(r)$ and effective pressure $p_{\rm eff}(r)$ decay sufficiently rapidly such that $p_{\rm eff}(r)r^2\to0$ and $m(r)\to\text{constant}$ as $r\to\infty$. It then follows from \ref{eqn:Nbrane2} that
\begin{equation}\label{eqn:Nbrinf}
    \mathcal{N}_{\rm brane}(r\to\infty)=2~.
\end{equation}
Accordingly, the requirement of stability necessitates that $\mathcal{N}_{\rm brane}(r)$ remain positive throughout the spacetime, thereby precluding the existence of circular null geodesics \cite{Hod:2024avc}.

Since the dominant energy condition applies to the physical matter variables $\rho$ and $p$, it need not hold for the effective quantities $\rho_{\rm eff}$ and $p_{\rm eff}$, which also receive contributions from the projected bulk Weyl tensor. Therefore, $\rho_{\rm eff}$ can be negative when the bulk contribution $\widetilde{U}$ is negative and dominates over the matter contribution, whereas $\rho_{\rm eff}$ is positive definite when the matter contribution is larger than the bulk contribution.  We therefore consider the following two cases.
\begin{itemize}
    \refstepcounter{casebrane} \label{case:brane1}
    \item \textit{Case 1}: If matter contribution dominates over bulk contribution, that is, $\rho_{\rm eff}\geq0$. We have
    \begin{equation}\label{eqn:case1}
    \rho\left(1+\frac{\rho}{2\lambda_{\rm b}}\right) \geq -3 \widetilde{U}~.
    \end{equation}
    Since $\rho_{\rm eff}\geq0$, from the definition of $m(r)$ in  \ref{eqn:brane-mass},  $m(r)\geq0$ for all values of $r$. Thus, using the condition in  \ref{eqn:maincondition} for the non-existence of photon rings, we find
    \begin{equation}\label{eqn:brane-ineq1}
    \begin{aligned}
     F_{\rm eff}<1~,
    \end{aligned}
    \end{equation}
where we have defined an effective force function that includes the effective radial pressure. Thus, it accounts for both the pressure due to the matter and the bulk contribution. Mathematically, $F_{\rm eff}$ is given by
\begin{equation}\label{eqn:brane-force}
    F_{\rm eff}=4\pi r^2 p_{\rm eff}~.
\end{equation}
Restoring the appropriate physical constants in four dimensions, the resulting upper bound takes the form
\begin{equation}\label{eqn:brane-ineq2}
    F_{\rm eff}<\frac{c^4}{G_4}~.
\end{equation}
Therefore, we see that the effective force bound is the same as the force bound obtained in the case of general relativity for $d=4$.\\
We now turn to deriving a bound on $F_{\rm brane}=4\pi r^2p$ which represents the force associated solely with the pressure due to the matter contribution. In the following analysis, we consider a positive brane tension, $\lambda_{\rm b}>0$, like in the case of the Randall–Sundrum II (RS-II) model \cite{Randall:1999vf}. We define a quantity $A=1-4\pi r^2\left(\widetilde{U}+2\widetilde{P}\right)~$. Depending on the sign of $A$, different energy conditions yield stronger bounds on the force. Specifically, the dominant energy condition provides a stronger bound for $A\geq0$ whereas the weak energy condition yields a stronger bound for $A<0$. Accordingly, we consider these two cases separately and further divide each case into two subcases as detailed below. The corresponding detailed derivations are presented in \ref{app:braneworld}.
    \begin{itemize}
        \renewcommand{\labelitemii}{$\circ$}
        \refstepcounter{subcasebrane} \label{subcase:brane1a}
        \item \textit{Case 1(a)}: $A\geq0$ \\  
        Using the inequality given in \ref{eqn:brane-ineq1}, the dominant energy condition stated in \ref{eqn:dominant} and the expression for $p_{\rm eff}$ given in \ref{eqn:effpress},  we deduce the general bound,
        \begin{equation}\label{eqn:brane-ineq3}
            F_{\rm brane}<A\left(1+\frac{3\rho}{2\lambda_{\rm b}}\right)^{-1}~.
        \end{equation}
        The detailed derivation of the bound is given in \ref{app:case1a}. Further, since $\rho\geq0$ and $\lambda_{\rm b}>0$, we can simplify the bound to
        \begin{equation}\label{eqn:brane-ineq4}
            F_{\rm brane}<A~.
        \end{equation}
        We infer that for $\widetilde{P}=0$ and $\widetilde{U}=0$, that is, in the absence of any higher-dimensional bulk contribution, the bound reduces to unity, which matches our results obtained in \ref{subsec:d-gr}, as the absence of bulk contributions recovers the four-dimensional general relativity case. Thus, the bound on the force from matter contribution depends on the bulk contributions $\widetilde{P}$ and $\widetilde{U}$. Depending on their sign and magnitude, the bound is greater or less than that of standard four-dimensional general relativity. Furthermore, if $\widetilde U$ and $\widetilde P$ scale as $\widetilde U \sim U_0/r^2$ and $\widetilde P \sim P_0/r^2$ respectively, the resulting upper bound becomes independent of the radial coordinate, yielding a radius-independent bound on $F_{\rm brane}$. \\

        \refstepcounter{subcasebrane} \label{subcase:brane1b}
        \item \textit{Case 1(b)}: $A<0$
        Proceeding like the previous subcase but instead using the weak energy condition implied by the dominant energy condition, $\rho+p\geq0$, we get the general bound in this case given by
        \begin{equation}\label{eqn:brane-ineq5}
            F_{\rm brane}<A\left(1+\frac{\rho}{2\lambda_{\rm b}}\right)^{-1}~.
        \end{equation}
        The corresponding derivation for this case is presented in \ref{app:case1b}. Unlike the previous case, the second term cannot be discarded to obtain a simplified bound because  $A<0$. Therefore, the bound in  \ref{eqn:brane-ineq5} must be retained in its present form. We note that in the limit $\widetilde{U}=\widetilde{P}=0$ and $\lambda_{\rm b}\to\infty$, the bulk-induced Weyl corrections and the quadratic brane corrections vanish respectively. Consequently, in this case too, the force bound reduces to unity as is obtained in $4$-dimensional general relativity \cite{Hod:2024avc}. 
    \end{itemize}

    \refstepcounter{casebrane} \label{case:brane2}
    \item \textit{Case 2}: If the bulk contribution dominates over the matter contribution, that is, $\rho_{\rm eff}<0$. The condition is given by
    \begin{equation}\label{eqn:case2}
    \rho\left(1+\frac{\rho}{2\lambda_{\rm b}}\right) < -3 \widetilde{U}~.
    \end{equation}
    Note that, unlike the previous cases, $m(r)<0$, as a result of which the bound on $F_{\rm eff}$ obtained in \autoref{case:brane1} is no longer valid in this case. Thus, unlike \autoref{case:brane1}, here we obtain a radial coordinate dependent bound for $F_{\rm eff}$ using \ref{eqn:maincondition} given by\\

    \begin{equation}\label{eqn:brane-ineq6}
    \begin{aligned}
     F_{\rm eff}<1-\frac{3m(r)}{r}~,
    \end{aligned}
    \end{equation}
    
    We now proceed to derive a bound on $F_{\rm brane}$. For convenience, we define $A'=1-3m(r)/r-4\pi r^2(\widetilde{U}+2\widetilde{P})$. As in the previous case, we divide the analysis into two subcases, depending on the sign of $A'$.

        \begin{itemize}
        \renewcommand{\labelitemii}{$\circ$}
        \refstepcounter{subcasebrane} \label{subcase:brane2a}
        \item \textit{Case 2(a)}: $A'\geq0$ \\
        Combining the inequality in \ref{eqn:brane-ineq6} with the dominant energy condition in \ref{eqn:dominant} and the expression for $p_{\rm eff}$ in \ref{eqn:effpress}, we obtain the general bound
      \begin{equation}\label{eqn:brane-ineq7}
           F_{\rm brane}<A'\left(1+\frac{3\rho}{2\lambda_{\rm b}}\right)^{-1}~.
      \end{equation}
      Furthermore, since $\rho\geq0$ and $\lambda_{\rm b}>0$, for $A\geq0$, this reduces the bound to
      \begin{equation}\label{eqn:brane-ineq8}
           F_{\rm brane}<A'~.
      \end{equation}
      The result is analogous to that obtained in \autoref{subcase:brane1a} with $A$ replaced by $A'$. In this case, an additional term $-3m/r$ appears in the bound. Since $m(r)<0$, no lower bound on the mass function can be imposed based solely on this condition. Consequently, this case does not reduce to the standard four-dimensional general relativistic limit in which the mass function is non-negative under the dominant energy condition.
      
        \refstepcounter{subcasebrane} \label{subcase:brane2b}
        \item \textit{Case 2(b)}: $A'<0$
        Following the same approach as in \autoref{subcase:brane1b} and employing the weak energy condition implied by the dominant energy condition, $\rho+p\geq0$, we obtain the bound
        \begin{equation}\label{eqn:brane-ineq9}
            F_{\rm brane}<A'\left(1+\frac{\rho}{2\lambda_{\rm b}}\right)^{-1}~.
        \end{equation}
        Using the condition in \ref{eqn:brane2}, we obtain the simplified bound
        \begin{equation} \label{eqn:brane-ineq10}
            F_{\rm brane}<-\frac{\rho A'}{3\widetilde{U}}~.
        \end{equation}
        The detailed analysis leading to this bound is presented in \ref{app:case2b}.
        \end{itemize}
\end{itemize}     
\subsection{{Pure Lovelock gravity}}
\label{subsec:Lovelock}
Having examined the effect of extra dimensions on the force bound, we now investigate the bound in higher-curvature theories of gravity, focusing on pure Lovelock gravity\cite{Lovelock:1971yv}. We restrict our analysis to asymptotically flat spacetimes, for which $\nu(r)\to 0$ and $\lambda(r)\to 0$ as $r\to\infty$. 
The Lagrangian density for pure Lovelock gravity of order $N$ \cite{Garraffo:2008hu, Cai:2006pq} is given by
\begin{equation}
\label{eqn:Lovelock-lagrangian}
\mathcal{L}^{(N)}_{\mathrm{Lovelock}}
= \sqrt{-g}\,\alpha_N\mathcal{R}^{(N)}~,
\end{equation}
where $\alpha_N$ is the coupling constant and
\begin{equation} \label{eqn:LovelockRN}
    \mathcal{R}^{(N)}
    = \frac{1}{2^N}
    \delta^{a_1 b_1 \cdots a_N b_N}_{c_1 d_1 \cdots c_N d_N}
    \mathcal{R}^{c_1 d_1}_{a_1 b_1}
    \cdots
    \mathcal{R}^{c_N d_N}_{a_N b_N}
\end{equation}
is the $N$-th order Lovelock invariant, and $g$ denotes the determinant of the metric tensor. The action for pure Lovelock gravity of order $N$ is given by
\begin{equation}
\label{eqn:Lovelock-action}
S_{\mathrm{Lovelock}}^{(N)}
=\frac{c^4}{16\pi G_d}
\int d^d x\,\mathcal{L}_{\mathrm{Lovelock}}^{(N)}~.
\end{equation}
Since the $N$-th order Lovelock invariant corresponds to the dimensional continuation of the Euler density in $2N$ dimensions, it contributes nontrivially to the local field equations only for $d > 2N$ \cite{Padmanabhan:2013xyr}. Thus, the allowed values of $N$ satisfy
\begin{equation} \label{eqn:Lovelock-general-condition}
    N \leq \left\lfloor \frac{d-1}{2} \right\rfloor.
\end{equation}

In particular, for odd and even spacetime dimensions, respectively, this condition reads
\begin{equation}\label{eqn:Lovelock-condition}
  \begin{aligned}
     N \leq \frac{d-1}{2}~  ~~\text{~~ where~$d$~is~odd~dimension }~,\\
     N \leq \frac{d-2}{2}~  ~~\text{~~ where~$d$~is~even~dimension }~.
  \end{aligned}  
\end{equation}
For a static, spherically symmetric spacetime sourced by an anisotropic fluid, described by the metric functions $\nu(r)$ and $\lambda(r)$, the field equations take the form \cite{Dadhich:2016wtb, Chakraborty:2021dmu}
\begin{equation}\label{eqn:Lovelock-1}
    8\pi \rho(r)=\frac{(1-e^{-\lambda})^{N-1}}{2^{N-1}r^{2N}}\left[rN \lambda 'e^{-\lambda}+(d-2N-1)(1-e^{-\lambda})\right]~,
\end{equation}
\begin{equation}\label{eqn:Lovelock-2}
    8\pi p(r)=\frac{(1-e^{-\lambda})^{N-1}}{2^{N-1}r^{2N}}\left[rN \nu 'e^{-\lambda}-(d-2N-1)(1-e^{-\lambda})\right]~.
\end{equation}
Integrating \ref{eqn:Lovelock-1}, we obtain a solution for $e^{-\lambda}$ given by
\begin{equation}\label{eqn:Lovelock-lambda}
   e^{-\lambda}=1-2 \left(\frac{m(r)}{r^{d-2N-1}}\right)^{1/N}~.
\end{equation}
The mass function is defined as
\begin{equation}
    \label{eqn:Lovelock-mass}
    m(r)=4\pi\int_0^{r}\rho(r')\,r'^{d-2}\,dr'.
\end{equation}
Imposing the horizonless condition in \ref{eqn:horizonless} on \ref{eqn:Lovelock-lambda}, we obtain
\begin{equation}
    \label{eqn:Lovelock-horizonless-condition}
    2^N m(r)<r^{d-2N-1}.
\end{equation}

Following the approach adopted in the preceding cases, we introduce a function $\mathcal{N}_{\rm Lovelock}(r)$ to characterize the existence of photon rings imposing the photon orbit condition $r\nu'=2$ in \ref{eqn:Lovelock-2} \cite{Chakraborty:2021dmu}. The photon ring condition is given by $\mathcal{N}_{\rm Lovelock}(r_{\rm ph})=0$, where $r_{\rm ph}$
denotes the location of the photon ring. The function takes the form
\begin{equation}\label{eqn:Lovelock-N1}
    \mathcal{N}_{\rm Lovelock}=(d-1)e^{-\lambda}-(d-2N-1)-8\pi \frac{2^{N-1}r^{2N}p(r)}{(1-e^{- \lambda})^{N-1}}~.
\end{equation}
Substituting \ref{eqn:Lovelock-lambda} into \ref{eqn:Lovelock-N1}, we express $\mathcal{N}_{\rm Lovelock}$ entirely in terms of the mass function $m(r)$ and the pressure $p(r)$ given by 
\begin{equation}
\label{eqn:Lovelock-N2}
\mathcal{N}_{\rm Lovelock}(r)
=2(1-d)\left(\frac{m(r)}{r^{d-2N-1}}\right)^{1/N}
+2N
-8\pi r^{2N}p(r)
\left(\frac{r^{d-2N-1}}{m(r)}\right)^{\frac{N-1}{N}}.
\end{equation}
In the asymptotic limit, we assume that $m(r)$ approaches a finite constant and that the radial pressure satisfies $p(r)r^{2+\frac{(N-1)(d-1)}{N}}\to 0$ as $r\to\infty$. Under these asymptotic conditions,
\begin{equation} \label{eqn:NLovelock-inf}
    \mathcal{N}_{\rm Lovelock}(r\to\infty)=2N~.
\end{equation}

Thus, the absence of circular null geodesics, a necessary condition for stability, requires $\mathcal{N}_{\mathrm{Lovelock}}(r)>0$ throughout the spacetime \cite{Hod:2024avc}.

As in the previous cases, we adopt the force definition $F_{\rm Lovelock}(r)=4\pi r^{d-2}p(r)$. Imposing the condition on $\mathcal{N}_{\rm Lovelock}$,
\begin{equation}\label{eqn:Lovelock-ineq1}
    \begin{aligned}
        &F_{\rm Lovelock}<\frac{Nm(r)^{1-1/N}}{r^{(1-1/N)(d-2N-1)-(d-2N-2)}}-\frac{(d-1)m(r)}{r}~.
    \end{aligned}
\end{equation}

The dominant energy condition, consistent with the assumptions adopted in the preceding cases, implies that the mass function defined in \ref{eqn:Lovelock-mass} satisfies $m(r)\geq 0$. Using this result, the inequality in ~\ref{eqn:Lovelock-ineq1} simplifies to
\begin{equation}
\label{eqn:Lovelock-ineq2}
    F_{\mathrm{Lovelock}}<N\frac{m(r)^{1-\frac{1}{N}}}{r^{3-\frac{d-1}{N}}}~.
\end{equation}

We now consider several cases corresponding to specific relations between $d$ and $N$, followed by the general case.

\begin{itemize}
    \refstepcounter{caseLovelock}
    \label{case:Lovelock1}
    \item \textit{Case 1}: $d=3N+1$

    Since $m(r)\leq \mathcal{M}$, where $\mathcal{M}$ denotes the ADM mass of the spacetime at infinity, the above inequality given by \ref{eqn:Lovelock-ineq2} can be expressed as
    \begin{equation} \label{eqn:Lovelock-ineq3}
        F_{\mathrm{Lovelock}}<N\frac{\mathcal{M}^{1-\frac{1}{N}}}{r^{3-\frac{d-1}{N}}}~.
    \end{equation}
    Substituting $d=3N+1$ into \ref{eqn:Lovelock-ineq4}, we obtain
    \begin{equation} \label{eqn:Lovelock-ineq4}
        F_{\mathrm{Lovelock}} < N \mathcal{M}^{1-\frac{1}{N}}~.
    \end{equation}
Consequently, an $r$-independent upper bound on the force is obtained for this case. 

\refstepcounter{caseLovelock}
\label{case:Lovelock2}
\item \textit{Case 2:} $d=2N+2$ \\
This corresponds to the maximum allowed value of $N$ in even spacetime dimensions as specified in \ref{eqn:Lovelock-condition}. Applying the horizonless condition from \ref{eqn:Lovelock-horizonless-condition} to Eq.~\ref{eqn:Lovelock-ineq2}, we obtain
\begin{equation}
\label{eqn:Lovelock-ineq5}
    F_{\mathrm{Lovelock}}<\frac{N}{2^{N-1}}r^{d-2N-2}~.
\end{equation}
Setting $d=2N+2$ eliminates the radial dependence from the bound, yielding
\begin{equation}
\label{eqn:Lovelock-ineq6}
    F_{\mathrm{Lovelock}}<\frac{N}{2^{N-1}}~.
\end{equation}
Hence, this case also admits an $r$-independent upper bound on the force.

\refstepcounter{caseLovelock}
\label{case:Lovelock3}
\item \textit{Case 3:} We now consider the general case for arbitrary $d$ and $N$. \\
Combining the bounds obtained in \ref{eqn:Lovelock-ineq3} and \ref{eqn:Lovelock-ineq5}, we obtain
\begin{equation} \label{eqn:Lovelock-ineq7}
F_{\mathrm{Lovelock}}<\min\left\{\frac{N\mathcal{M}^{1-\frac{1}{N}}}{r^{3-\frac{d-1}{N}}},\frac{N}{2^{N-1}}r^{d-2N-2}\right\}~.
\end{equation}
We now introduce a rescaled force $\mathcal{F}_{\mathrm{Lovelock}}$, with the rescaling chosen according to which of the two terms in \ref{eqn:Lovelock-ineq7} provides the tighter bound. When the first term is the minimum, we define
\begin{equation} \label{eqn:Lovelock-general-rescale1}
\mathcal{F}_{\mathrm{Lovelock}}=\frac{F_{\mathrm{Lovelock}}}{r^{\frac{d-1}{N}-3}}~.
\end{equation}
which gives
\begin{equation} \label{Lovelock-ineq8}
\mathcal{F}_{\mathrm{Lovelock}}<N\mathcal{M}^{1-\frac{1}{N}}.
\end{equation}
Alternatively, when the second term is the minimum, we define
\begin{equation} \label{eqn:Lovelock-general-rescale2}
\mathcal{F}_{\mathrm{Lovelock}}=\frac{F_{\mathrm{Lovelock}}}{r^{d-2N-2}}~,
\end{equation}
yielding
\begin{equation} \label{Lovelock-ineq9}
\mathcal{F}_{\mathrm{Lovelock}}<\frac{N}{2^{N-1}}~.
\end{equation}
\end{itemize}
Depending on the case considered, the bound is multiplied by the expression in \ref{eqn:dimanalysis} to restore the appropriate physical units. \par
For $d=4$ and $N=1$, both \autoref{case:Lovelock1} and \autoref{case:Lovelock2} apply, and in both cases, we obtain a force bound of unity consistent with the standard $4$-dimensional general relativistic result \cite{Hod:2024avc}. For $d$-dimensional general relativity, corresponding to $N=1$, we consider \autoref{case:Lovelock3}. In this limit, the two expressions in the minimum in \ref{eqn:Lovelock-ineq7} coincide, yielding the same bound. The rescaled force in this case obtained from the definitions in \autoref{case:Lovelock3} is given by
\begin{equation}
\mathcal{F}_{\mathrm{GR},d}=\frac{F_{\mathrm{GR},d}}{r^{d-4}}~,
\end{equation}
which satisfies
\begin{equation}
\mathcal{F}_{\mathrm{GR},d}<1,
\end{equation}
which is in agreement with the result obtained in \ref{subsec:d-gr}. 

\subsection{{Einstein-Gauss-Bonnet gravity}}
\label{subsec:EGB}
After investigating the force bound in pure Lovelock gravity in the preceding section, we now consider five-dimensional Einstein-Gauss–Bonnet (EGB) gravity, a particular case of the general Lovelock framework. This theory extends the standard Einstein–Hilbert action by including the second-order Lovelock contribution, providing a useful setting in which to investigate how the force bound is modified by higher-curvature terms.

The Lagrangian density for Einstein–Gauss–Bonnet gravity is given by
\begin{equation}\label{eqn:EGB-lagrangian}
    \mathcal{L}_{\rm EGB}=\sqrt{-g}(\alpha_0+\alpha_1R+\alpha_2 {L}_{\rm GB})~,
\end{equation}
where
\begin{equation}\label{eqn:EGB-lag-term}
    {L}_{\rm GB}=R^2+R_{\alpha \beta \mu \nu}R^{\alpha \beta \mu \nu}-4R_{\mu \nu}R^{\mu \nu}~.
\end{equation}
For an asymptotically flat spacetime, we set the cosmological constant contribution, represented by $\alpha_0$, to zero. Furthermore, adopting the usual normalization of the Einstein–Hilbert term, the action becomes
\begin{equation}\label{eqn:EGB-action}
    S_{\rm EGB}=\frac{1}{16 \pi G_5}\int d^5x\sqrt{-g}(R+\alpha_2 L_{\rm GB})~,
\end{equation}
where $\alpha_2$ denotes the Gauss–Bonnet coupling constant, which determines the relative strength of the higher-curvature contribution. In the limit $\alpha_2\rightarrow 0$, the Gauss–Bonnet term vanishes, and the theory reduces to general relativity.

Considering an anisotropic perfect fluid as the matter source, the gravitational field equations in Einstein–Gauss–Bonnet gravity can be written as \cite{Bogadi:2023moe, Pretel:2021gqq, Chakraborty:2021dmu}
\begin{equation}\label{eqn:EGB-fieldeqn1}
    8 \pi r^2 \rho(r)=r \lambda'e^{- \lambda}+2(1-e^{-\lambda})+2\alpha_2\lambda'e^{-\lambda} \frac{(1-e^{-\lambda})}{r}~ ,
\end{equation}
and
\begin{equation}\label{eqn:EGB-fieldeqn2}
    8 \pi r^2 p(r)=r \nu'e^{- \lambda}-2(1-e^{-\lambda})+2\alpha_2\nu'e^{- \lambda} \frac{(1-e^{-\lambda})}{r} ~.
\end{equation}
These equations involve the two metric functions, $\lambda(r)$ and $\nu(r)$. Integrating \ref{eqn:EGB-fieldeqn1} yields the following expression for $e^{-\lambda}$
\begin{equation}\label{eqn:EGB-lambda}
    e^{-\lambda}=1+\frac{r^2}{2 \alpha_2}-\frac{r^2}{2 \alpha_2} \sqrt{1+\frac{8 \alpha_2 m(r)}{r^4}}~.
\end{equation}
Here, the mass function \(m(r)\) is defined as
\begin{equation}\label{eqn:EGB-mass}
    m(r)=4\pi\int_0^{r}dr'\rho(r')r'^3~.
\end{equation}

Using the algebraic expression for the photon circular orbit, that is, $r \nu'=2$ in the pressure equation given by \ref{eqn:EGB-fieldeqn2}, we define a function $\mathcal{N}_{\rm EGB}(r)$ as in previous sections such that $\mathcal{N}_{\rm EBG}(r_{\rm ph})=0$, where $r_{\rm ph}$ is the radial coordinate for the circular photon orbit \cite{Chakraborty:2021dmu}. $\mathcal{N}_{\rm EGB}(r)$ is given by,
\begin{equation}\label{eqn:NEGB-1}
    \mathcal{N}_{\rm EGB}(r)=4 e^{- \lambda}-2-8 \pi r^2p(r)+\frac{4 \alpha_2}{r^2}e^{-\lambda}(1-e^{-\lambda})~.
\end{equation}
Substituting the expression for $e^{-\lambda}$ given by \ref{eqn:EGB-lambda} in the above expression for $\mathcal{N}_{EBG}$ we obtain,
\begin{equation}\label{eqn:NEGB-2}
    \mathcal{N}_{\rm EGB}(r)=-8 \pi r^2p+2\sqrt{1+\frac{8 \alpha_2 m}{r^4}}-\frac{8m}{r^2}.
\end{equation}
In the asymptotic regime, we assume that $m(r)$ approaches a finite constant and that the radial pressure satisfies $p(r)r^2\to 0$ as
$r\to\infty$. Under these conditions, we obtain
\begin{equation} \label{eqn:NEGB-inf}
\mathcal{N}_{\rm EGB}(r\to\infty)=2~.
\end{equation}
Since the absence of circular null geodesics is a necessary condition for stability, we require $\mathcal{N}_{\mathrm{EGB}}(r)>0$ throughout the spacetime \cite{Hod:2024avc}. Using this condition,
\begin{equation}\label{eqn:EGB-ineq1}
   \begin{aligned}
    &4 \pi r^3 p<r \sqrt{1+\frac{8 \alpha_2 m}{r^4}}-\frac{4m}{r}~.
  \end{aligned}
\end{equation}
Under the dominant energy condition, $\rho\geq 0$, which implies that the mass function satisfies $m(r)\geq 0$. Substituting this condition into  \ref{eqn:EGB-ineq1}, we obtain the simplified inequality
\begin{equation} \label{eqn:EGB-ineq2}
4\pi r^3 p<r\sqrt{1+\frac{8\alpha_2 m}{r^4}}~.
\end{equation}
Defining the force as $F_{\rm EGB}=\Omega_3 r^3p$, \ref{eqn:EGB-ineq2} can be expressed as
\begin{equation} \label{eqn:EGB-ineq3}
F_{\rm EGB}<r\sqrt{1+\frac{8\alpha_2 m}{r^4}}~.
\end{equation}
For $d=5$, the rescaled force is defined as $\mathcal{F}_{\rm EGB}=F_{\rm EGB}/r^{d-4}=F_{\rm EGB}/r$. Consequently, \ref{eqn:EGB-ineq3} yields the bound
\begin{equation} \label{eqn:EGB-ineq4}
\mathcal{F}_{\rm EGB}<\sqrt{1+\frac{8\alpha_2 m}{r^4}}~.
\end{equation}
For a negative Gauss--Bonnet coupling \cite{Wright:2015yda, Dehghani:2004cf}, $\alpha_2\leq0$, the condition $m(r)\geq 0$ ensures that the square-root factor does not exceed unity. We therefore obtain the definite upper bound in this case 
\begin{equation} \label{eqn:EGB-ineq5}
\mathcal{F}_{\rm EGB}<1~.
\end{equation}
Restoring the appropriate physical dimensions, this bound takes the form
\begin{equation} \label{eqn:EGB-ineq6}
\mathcal{F}_{\rm EGB}<\frac{c^4}{G_5}~,
\end{equation}
where $G_5$ denotes the five-dimensional gravitational constant.

\section{\textbf{Conclusion}}
\label{sec:conclusion}

The possibility that nature may host horizonless objects, including potential black-hole mimickers, has attracted considerable attention, although their physical nature and composition remain incompletely understood. Investigating the conditions under which such configurations can exist without developing event horizons is therefore of considerable interest. In this context, force bounds provide a useful framework for constraining pressure-supported configurations and examining the role of gravitational dynamics in determining their physical properties.

In this work, we have investigated the existence and formulation of force bounds for horizonless, spatially regular configurations supported by an anisotropic fluid in several gravitational theories, namely, $d$-dimensional general relativity, braneworld gravity, pure Lovelock gravity, and five-dimensional Einstein-Gauss-Bonnet gravity. Our objective has been to determine bounds on the force or appropriately rescaled force and to examine how additional dimensions, bulk contributions, and higher-curvature corrections influence these bounds.

Within $d$-dimensional general relativity, we obtain a bound on the rescaled force. In the braneworld scenario, the bounds depend on the bulk Weyl contributions represented by the dark radiation $\widetilde{U}$ and the dark pressure $\widetilde{P}$. They also depend on the sign of the effective energy density and on the relative contributions of matter and the bulk. Thus, the force bound associated with the pressure arising from matter contributions can differ from that obtained in four-dimensional general relativity.

In pure Lovelock gravity, we derive a bound on the force for a specific relation between the spacetime dimension $d$ and the Lovelock order $N$. By introducing an appropriately rescaled force, we further obtain a bound for the general case considered in our analysis. This demonstrates the importance of accounting for dimensional dependence when investigating maximal-force bounds beyond general relativity.

Finally, in five-dimensional Einstein--Gauss--Bonnet gravity, we derive a bound on the rescaled force. We show that for a negative Gauss--Bonnet coupling, $\alpha<0$, it reduces to an $r$-independent bound under the conditions considered in our analysis. This result illustrates how higher-curvature corrections can affect the existence and form of force bounds and emphasizes that their applicability depends on the specific gravitational dynamics.

The force bounds derived in this work may have several potential applications in the study of compact objects and gravitational physics. If these theoretical bounds can be related to observable quantities and tested against astrophysical data, they may help constrain the properties of horizonless compact objects and assess the viability of different gravitational theories. In particular, they could be used to investigate the range of pressure distributions compatible with horizonless configurations and to constrain their possible compactness within different gravitational theories. An interesting direction for future work would be to investigate whether these bounds can be related to the existence or absence of photon spheres, which influence the gravitational lensing properties and characteristic oscillation spectra of compact objects. Such an investigation could clarify whether the force bounds impose additional restrictions on the properties of these configurations. Furthermore, comparing the bounds obtained in general relativity with those in braneworld and higher-curvature gravity may help identify differences between competing descriptions of strong-field gravity. Future studies could explore whether these differences lead to observable consequences in gravitational wave signals or other astrophysical probes \cite{Johnson-Mcdaniel:2018cdu, Cardoso:2016oxy, Cardoso:2019rvt}. Establishing such connections would require constructing explicit models of compact objects and determining their observable properties, which we leave for future investigation.

Taken together, our findings indicate that the existence and form of maximal-force bounds depend on the underlying gravitational theory, its dimensional structure and the physical conditions imposed on matter and spacetime geometry. Although the characteristic scale $c^4/G$ provides a natural reference for the maximal-force conjecture in four-dimensional general relativity, extending this notion to higher dimensions and modified theories of gravity requires careful consideration of the corresponding field equations and assumptions. The results therefore suggest that maximal-force bounds cannot be straightforwardly regarded as universal across all the gravitational theories considered here. 

\section*{Acknowledgements}
The author would like to thank Sumanta Chakraborty and Ribhu Paul for valuable discussions and comments on the manuscript. The author also acknowledges the Institute of Mathematical Sciences for financial support.
\newcounter{supmat}
\renewcommand{\thesupmat}{\Roman{supmat}}

\section*{Appendix}
\appendix
\labelformat{section}{Appendix #1} 
\labelformat{subsection}{Appendix #1}

\section{Force bounds in braneworld gravity}
\label{app:braneworld}
In this appendix, we present the detailed derivations of the force bounds in braneworld gravity for the different cases discussed in \ref{subsec:braneworld}. We begin with the case where $\rho_{\rm eff}\geq 0$ and derive bounds on the force associated with the pressure due to the matter contribution under different conditions on the respective quantities $A$ and $A'$. The derivations employ the non-existence condition for circular null geodesics together with the appropriate energy conditions.

\subsection{Force bound for \texorpdfstring{$\rho_{\rm eff}\geq 0$ and $A\geq 0$}{rho\_eff >= 0 and A' >= 0}}
\label{app:case1a}
Using the inequality in \ref{eqn:brane-ineq1} together with the expression for $p_{\rm eff}$ given in \ref{eqn:effpress}, we obtain,

\begin{equation}\label{eqn:appendix1a-1}
\begin{aligned}
&4\pi r^2\left[p+\frac{\rho}{2 \lambda_{\rm b}}(\rho+2p)
+\widetilde{U}+2\widetilde{P}\right]<1~.
\end{aligned}
\end{equation}
The dominant energy condition given in Eq.~\ref{eqn:dominant} implies that $\rho\geq p$. Applying this inequality to the above expression, we obtain
\begin{equation}\label{eqn:appendix1a-2}
\begin{aligned}
&4\pi r^2\left[p+\frac{3p\rho}{2 \lambda_{\rm b}}
+\widetilde{U}+2\widetilde{P}\right]<1\\
\implies &F_{\rm brane}<A\left(1+\frac{3\rho}{2\lambda_{\rm b}}\right)^{-1}~.
\end{aligned}
\end{equation}

\subsection{Force bound for \texorpdfstring{$\rho_{\rm eff}\geq 0$ and $A<0$}{rho\_eff >= 0 and A' < 0}}
\label{app:case1b}

Since the dominant energy condition assumed in this work implies the weak energy condition, we have $\rho+p\geq0$. Combining this condition with \ref{eqn:appendix1a-1}, we obtain,
\begin{equation}\label{eqn:appendix1b-1}
    \begin{aligned}
    &F_{\rm brane}+4\pi r^2\left(\frac{\rho p}{2 \lambda_{\rm b}}\right)<1-8\pi r^2\widetilde{P}-4 \pi r^2\left[\frac{\rho}{2 \lambda_{\rm b}}(\rho +p)\right]-4\pi r^2\widetilde{U}\\
    \implies &F_{\rm brane}\left(1+\frac{\rho}{2\lambda_{\rm b}}\right)<A~\\
    \implies &F_{\rm brane}<A\left(1+\frac{\rho}{2\lambda_{\rm b}}\right)^{-1}~.
    \end{aligned}
\end{equation}

\subsection{Force bound for \texorpdfstring{$\rho_{\rm eff}<0$ and $A'<0$}{rho\_eff < 0 and A' < 0}}
\label{app:case2b}
From the condition in \ref{eqn:case2}, together with $\rho\geq 0$ and $\lambda_{\rm b}>0$, we infer that $\widetilde{U}<0$. Consequently, we obtain
\begin{equation} \label{eq:appendix2b-2}
    \left(1+\frac{\rho}{2\lambda_{\rm b}}\right)^{-1}>- \frac{\rho}{3\widetilde{U}}~.
\end{equation}
Multiplying both sides by $A'<0$ reverses the inequality, yielding,
\begin{equation} \label{eq:appendix2b-3}
    A'\left(1+\frac{\rho}{2\lambda_{\rm b}}\right)^{-1}<- \frac{A'\rho}{3\widetilde{U}}~.
\end{equation}
Combining this result with \ref{eqn:brane-ineq9}, we obtain,
\begin{equation} \label{eq:appendix2b-4}
    F_{\rm brane}<-\frac{\rho A'}{3\widetilde{U}}~.
\end{equation}
The above simplifications have been incorporated in the main text.
        

\bibliographystyle{utphys1}
\bibliography{mybib}

\end{document}